\documentclass[twocolumn,showpacs,prb,floatfix,superscriptaddress]{revtex4}

\usepackage{color} 
\usepackage{tabularx} 
\usepackage{epsfig}
\usepackage{amsmath} 
\usepackage{amssymb} 
\usepackage{graphicx}
\usepackage{wasysym}

\begin{document}

\title{Sampling the ground-state magnetization of
$d$-dimensional $p$-body Ising models}

\author{Creighton K.~Thomas}
\affiliation {Department of Physics and Astronomy, Texas A\&M University,
College Station, Texas 77843-4242, USA}

\author{Helmut G.~Katzgraber} 
\affiliation {Department of Physics and Astronomy, Texas A\&M University,
College Station, Texas 77843-4242, USA}
\affiliation {Theoretische Physik, ETH Zurich, CH-8093 Zurich, Switzerland}

\date{\today}

\begin{abstract}

We demonstrate that a recently introduced heuristic optimization
algorithm [Phys.~Rev.~E {\bf 83}, 046709 (2011)] that combines a local
search with triadic crossover genetic updates is capable of sampling
nearly uniformly among ground-state configurations in spin-glass-like
Hamiltonians with $p$-spin interactions in $d$ space dimensions that
have highly degenerate ground states.  Using this algorithm we probe the
zero-temperature ferromagnet to spin-glass transition point $q_c$ of
two example models, the disordered version of the two-dimensional
three-spin Baxter-Wu model [$q_c = 0.1072(1)$] and the three-dimensional
Edwards-Anderson model [$q_c = 0.2253(7)$], by computing the Binder
ratio of the ground-state magnetization.

\end{abstract}

\pacs{75.50.Lk, 75.40.Mg, 05.50.+q, 64.60.-i}

\maketitle

\section{Introduction}

Disordered systems with degenerate ground states often have an
exponentially-large\cite{comment:N} number of states that minimize the
Hamiltonian (cost function) of the problem.  A hallmark model system to
study degenerate ground states is the Edwards Anderson Ising spin-glass
model with bimodal ($\pm J$) interactions,\cite{binder:86} for which the
entropy is extensive even at zero temperature due to the exponential
number of ground-state configurations.

Studying these systems at finite temperatures using Monte Carlo
simulations typically poses no major challenges: An estimate of an
observable (e.g., the magnetization) has to be measured and averaged
over Monte Carlo time. This means that an average over different
degenerate states is automatically taken into account. However, when
studying the ground-state properties of such systems, it is imperative
to ensure that an average over either all ground-state configurations or
at least an unbiased subset is taken. While some algorithms are known to
do a relatively fair sampling of the ground-state configurations (e.g.,
simulated annealing),\cite{kirkpatrick:83,moreno:03} others, such as
quantum annealing,\cite{finnila:94,santoro:02,das:05,matsuda:09} have
been shown to favor certain configurations and exponentially suppress
others.

Although finding ground-state energies is not typically harder with
discrete energy distributions than in models with continuous energies,
determining thermodynamic quantities requires thorough checks of the
algorithms used.  Numerically exact results are achievable only in
limited scenarios,\cite{klotz:94,poulter:05,thomas:11c} though
comparisons with theoretical predictions also may provide confidence
that the technique is behaving
correctly.\cite{boettcher:04,boettcher:05}

Here, we investigate the effectiveness of a heuristic optimization
algorithm previously developed\cite{thomas:11} for finding all ground
states of a system, or, if that number is too large, finding a {\em
representative and unbiased} sample, applied to a $p$-spin
generalization of the $d$-dimensional Edwards-Anderson Ising spin-glass
model.\cite{comment:p2,edwards:75} The general applicability of this
algorithm implies that it is useful for studying a wide range of hard
problems where it is desirable to average over {\em many} degenerate
optimal cases.  We illustrate the usefulness of the method for studying
spin-glass models by computing the zero-temperature
ferromagnet--to--spin-glass phase transition as the disorder strength is
varied in the disordered two-dimensional three-spin Baxter-Wu model and
in the three-dimensional two-spin Edwards-Anderson model with bimodal
interactions.

Using a moderate numerical effort, we show for the case of the
three-dimensional two-spin Edwards-Anderson Ising spin-glass model that
the critical threshold $q_c$ between a ferromagnetic and a spin-glass
phase is $q_c = 0.2253(7)$, in agreement with previous
studies,\cite{hartmann:99b} albeit with considerably smaller error bars.

In Sec.~\ref{sec:num} we present an overview of the algorithm,
followed by results on the two-dimensional three-spin model in
Sec.~\ref{sec:3sp} and results on the three-dimensional Edwards-Anderson
spin glass in Sec.~\ref{sec:3d}.

\section{Numerical Method}
\label{sec:num}

The Edwards Anderson model in three space dimensions has been shown to
be NP-hard,\cite{barahona:82} as has the spin glass with three-body
interactions in two dimensions.\cite{thomas:11} It is therefore expected
that there are no exact algorithms capable of simulating large instances
of either model at zero temperature. The genetic algorithm presented in
Ref.~\onlinecite{thomas:11} heuristically generates many optimal
ground-state configurations of disordered $p$-spin models.  The method
starts from a population of $N_p$ random spin configurations and
combines a highly-effective local search heuristic with triadic
crossover genetic updates.\cite{pal:94} Although very large systems are
unattainable with such an approach, it was shown to produce ground-state
energies in the two-dimensional three-spin model with high confidence
for $N \lesssim 42^2 = 1764$ spins.\cite{thomas:11}

\paragraph*{Genetic algorithm:} The algorithm proceeds
according to the following steps

\begin{enumerate}

\item Initialize $N_p$ configurations: for each configuration

\begin{enumerate}

    \item Initialize the spins to a high-temperature configuration where
          all spins are assigned randomly.

    \item Perform a local search update to generate a low-energy
	  configuration.

    \item If this configuration is already present in the population,
	  go back to 1(a).

    \item Otherwise, accept this configuration and proceed to the next one.

\end{enumerate}

\item Perform a triadic crossover update to generate two new
      configurations, $C_1$ and $C_2$.

\begin{enumerate}

    \item Select three parents $P_1$, $P_2$ and $P_3$ randomly.

    \item For each spin $i$

	\begin{enumerate}

	    \item If $P_1(i) = P_3(i)$, $C_1(i) \leftarrow P_1(i)$
	    and $C_2(i) \leftarrow P_2(i)$

	    \item Otherwise $C_1(i) \leftarrow P_2(i)$ and $C_2(i)
	    \leftarrow P_2(i)$

	\end{enumerate}

    \item Select two configurations $X$ and $Y$ from the population
    at random.

    \item If $E(C_1) < E(X)$, replace $X$ with $C_1$; if $E(C_2) <
     E(Y)$ replace $Y$ with $C_2$.

    \end{enumerate}

\item Repeat step 2 $N_\mathrm{steps}$ times or until all states of
      the system have the same energy.

\end{enumerate}
Unlike in many genetic algorithms, mutations are not necessary to
achieve good results with this method.

\paragraph*{Local search component:} The local search algorithm used in step
1(b) above (described in detail in Ref.~\onlinecite{thomas:11})
generates clusters of nearby spins for which the energy is lowered
if all spins in the cluster are flipped.  This is achieved by
performing a depth-first search starting from each spin in the
system. The search algorithm takes two parameters, $E_\mathrm{max}\geq 0$
and $d_\mathrm{max}$.  At each step of the search, the spins adjacent
to the current site that are not already in the cluster are considered to be
added.  As each
spin is added, the energy of flipping the current spin and all its
ancestors in the search tree (all current members of the cluster)
is maintained.  If the energy is negative, the algorithm has found
an energy-lowering move, so all spins in the current cluster are
flipped and the search terminates successfully.  If the energy
exceeds the threshold energy $E_\mathrm{max}$ or the search depth
becomes more than $d_\mathrm{max}$, this search branch is discarded.
For $E_\mathrm{max}$ and $d_\mathrm{max}$ both large, this search
algorithm gives strong results, but runs slowly.  For use as part
of the genetic algorithm above, speed is more important than finding
the lowest energy, therefore we search for small clusters with few
barriers, setting $E_\mathrm{max}=4$ (the smallest possible change
in energy in our model) and $d_\mathrm{max} = L$, where $L$ is the
linear size of the system.

With $N_p$ and $N_\mathrm{steps}$ sufficiently large, this genetic
algorithm produces ground states with high confidence.\cite{thomas:11}
In general, to study a ferromagnet--paramagnet transition, it is useful
to study the finite-size scaling of dimensionless quantities based on
the magnetization of the system.  By computing quantities such as the
Binder ratio,\cite{binder:81} one can straightforwardly measure the
location of the critical point, as well as the critical exponent $\nu$.
When the disorder distribution is continuous, each disorder sample has
a unique ground state (up to global degeneracies), so finding this
ground-state configuration immediately yields both the energy and
magnetization; computing its energy and magnetization are therefore
typically of comparable difficulty.  But in the discrete-disorder
case, the number of ground-state configurations $N_g$ is typically
very large.  Computing the magnetization requires either finding
all ground states, or generating a uniform sampling to find a subset
of these.  This heuristic algorithm does not stop after it finds one
ground-state configuration; it continues finding others until either
all ground states are found or the algorithm can find no more ground
states due to a built-in constraint.

\begin{figure}[tb]
\includegraphics[width=\columnwidth]{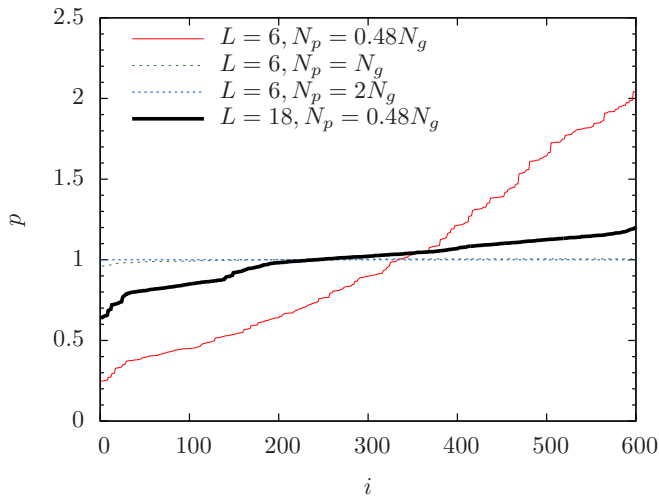}
\caption{(Color online) 
Normalized relative probability $p$ of each ground state occurring for
varying population sizes.  A single instance of disorder with $N = 6\times 6$
spins is considered, in addition to one instance of disorder with $N= 18\times
18$.  The disorder cases were selected, for comparison, 
to have the same number of ground states, with $N_g=600$.
The ground states (indexed by $i$)
are sorted according to their probability of occurring, giving a
monotonically increasing plot.  Even with $N_p < N_g$, the
probability for finding each ground state is a factor of order unity
times the uniform probability, implying that one can do a reasonable
approximation of the $T=0$ thermodynamic state even when $N_p < N_g$.
With other factors being equal, the larger system size shows a more uniform
sampling.
}
\label{fig:all_ground_states}
\end{figure}

For a genetic algorithm, the quality of the results depends both on the run
time of the algorithm and the number of configurations being concurrently
updated, the population size $N_p$.  For sufficient run time, our numerical
results show that this algorithm finds all the ground-state configurations of a
degenerate system if the number of ground states $N_g < N_p$.  When $N_g >
N_p$, which will inevitably be the case for some large systems, the
probabilities for finding each ground state vary somewhat, although our tests
suggest that the probability of finding a particular ground state is within
some factor times the expected probability with a uniform distribution, with
that factor typically being of order unity.  For one instance of the 3-spin
model studied in Sec.~\ref{sec:3sp} with 600 ground states, we reran our
algorithm $10^5$ times to find the relative probability for the genetic
algorithm to find each ground state.  Note that disorder instances with this
number of ground states are rare for $L=6$; the median number of ground states
we find over all disorder instances is 4.  For this sample, when $N_p = 2 N_g$
all ground states were found in every run, and for $N_p = N_g$ almost all
ground states were found in nearly every runs.  When $N_p = 0.48 N_g$, the
ground state with the largest basin of attraction is roughly eight times more
likely to be seen than the ground state with the smallest.  To probe the system
size dependence, we also looked at a larger system of size $L=18$, where we
chose a sample that also has 600 ground states.  In this case, the algorithm
selects among those ground states much more uniformly, with only a factor of 2
between most and least likely ground states.  For $L=18$, the median number of
ground states is 768; with $N_p>5000$, a large majority of the disorder
instances considered have $N_g<N_p$ and are expected to be treated fairly by
the algorithm.  These results are shown in Fig.~\ref{fig:all_ground_states}.

Even when $N_g > N_p$, this performance strongly outperforms quantum
annealing\cite{finnila:94,santoro:02,das:05,das:08} where in the case of
the fully-frustrated Villain model\cite{villain:77} certain ground-state
configurations are exponentially suppressed.\cite{matsuda:09} We have
not found any trends in the magnetizations produced by the algorithm;
using only this magnetization data, the ground states appear to be
chosen in random order, even when $N_g > N_p$, as is shown in
Sec.~\ref{sec:3sp}, and Fig.~\ref{fig:magnetization}.  An average
over the configurations that have been found therefore appears to be
representative of the entire ensemble for this algorithm.

\section{Three-spin Ising Model}
\label{sec:3sp}

\subsection{Model}

The three-spin Ising model with random plaquette interactions is given
by the Hamiltonian
\begin{equation}
\mathcal{H} = - J \sum_{\bigtriangleup} 
                \eta_{\bigtriangleup}
                S_{\bigtriangleup}^1
                S_{\bigtriangleup}^2
                S_{\bigtriangleup}^3 , 
\end{equation}
where the sum is over all triangular plaquettes in a two-dimensional
triangular lattice and each term involves the product of the Ising
spins $S_{\bigtriangleup}^i \in \{\pm 1\}$ on the vertices of each
plaquette. Without any loss of generality we set $J \equiv 1$.  The
plaquette couplings $\eta_{\bigtriangleup}$ are chosen independently
and randomly according to a bimodal distribution, that is,
\begin{equation}
P(\eta_{\bigtriangleup}) =      (1-q)\delta(\eta_{\bigtriangleup}-1)
                                +q\delta(\eta_{\bigtriangleup}+1)
\label{eq:bimodal}
\end{equation}
with $q=0$  being the pure ferromagnetic case.  As $q$ increases,
the model goes through a phase transition from ferromagnet to
paramagnet.  A previous study at finite temperature using Monte Carlo methods
finds this transition to occur at $q_c =
0.109(2)$.\cite{katzgraber:09c}

This model is of interest for a number of reasons.  $p$-spin
models have been related to structural
glasses.\cite{kirkpatrick:87,kirkpatrick:87b,kirkpatrick:87c,larson:10} Furthermore,
when computing the error threshold of topological color
codes\cite{bombin:06,katzgraber:09c} to bit-flip errors the problem
maps\cite{dennis:02} onto the 3-spin model in two space dimensions.
Although the error threshold is given by the crossing of the
ferromagnetic--paramagnetic phase boundary line with the Nishimori
line,\cite{nishimori:81} the zero-temperature phase boundary delivers
a {\em lower bound} for the threshold that is often easier to compute
than to equilibrate finite-temperature Monte Carlo simulations.

The phase diagram for the Ising model on a square lattice shares
much in common with the three-spin Ising model on a triangular
lattice.  The three-spin Ising model with uniform ferromagnetic
interactions is exactly solvable and known as the Baxter-Wu
model,\cite{baxter:82} and the transition temperature on a triangular
lattice is identical to that of the two-dimensional Ising model on
a square lattice. Although the models are in different universality
classes,\cite{katzgraber:09c,yeomans:92} the two phase diagrams
are quite similar even in the presence of disorder.  Numerically,
it appears that the multicritical points for these two models
are very close to one another or identical.\cite{katzgraber:09c}
The results presented here suggest that this similarity does not
extend to the low-temperature regime {\em below} the Nishimori
line,\cite{nishimori:81} with possible reentrant behavior being much
weaker, if it exists at all, in this model.

In contrast to the Ising model with $p=2$, which has a two-fold
(global spin flip) degeneracy, this model is fourfold degenerate.
The magnetization
\begin{equation}
M = \frac{1}{N}\sum_i S_i
\end{equation}
in the ordered state is not symmetric about zero, although the average
over the four degeneracy sectors related by symmetry operations
is zero: in the pure ferromagnet, one of the four degenerate ground
states has $M=1$, while the other three states each have $M=-1/3$. In
the paramagnetic state there is no state which is much stronger than
the others, and the disorder-averaged magnetization in each of these
four sectors is zero.

The zero-temperature thermodynamic state consists
of all ground states with equal probability. To compute properties
related to the magnetization, it is therefore {\em not sufficient}
to find an arbitrary specific ground state of the model unless it
can be demonstrated that it is chosen in an unbiased way.

\subsection{Sampling results}

When $N_p > N_g$, it is seen that the number of ground states found
for the three-spin model is always a multiple of $4$, and the average
magnetization is $0$. It is necessary that these both be true due to
the fourfold degeneracy of the system. This result suggests that the
algorithm is finding all ground states of the system.  It is possible
that there are other states which are not found, but we emphasize
that the algorithm is given no information about the symmetries of the
problem, so states which are related to one another by global spin-flip
symmetries are not necessarily easy for the algorithm to link together.
Thus is appears that we are seeing all ground states when $N_p > N_g$.

\begin{figure}[tb]
\includegraphics[width=\columnwidth]{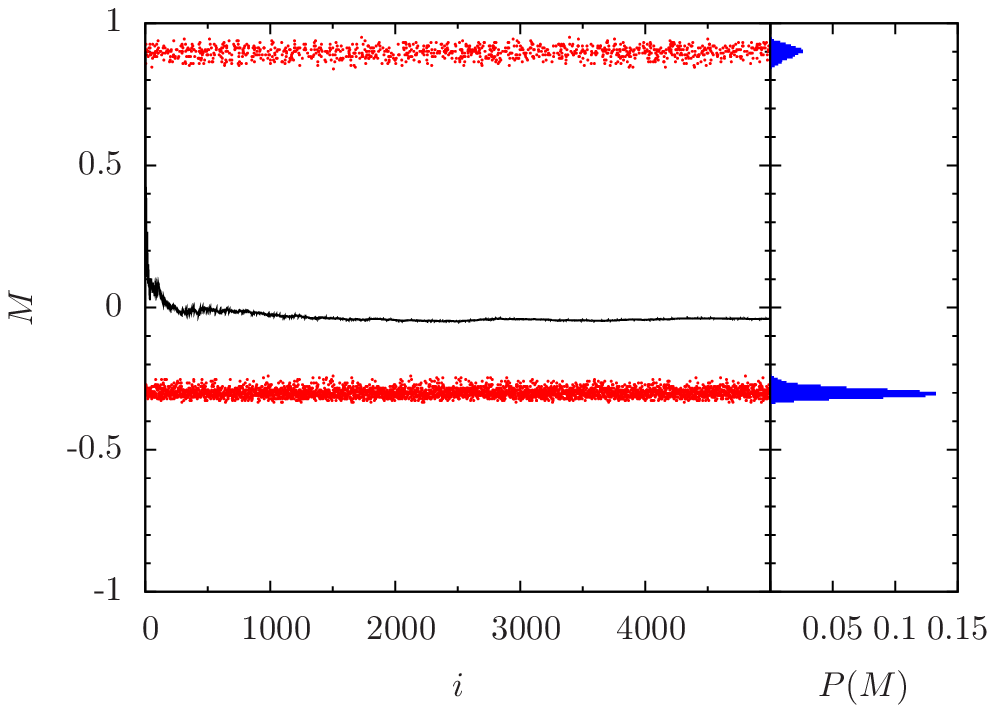}
\includegraphics[width=\columnwidth]{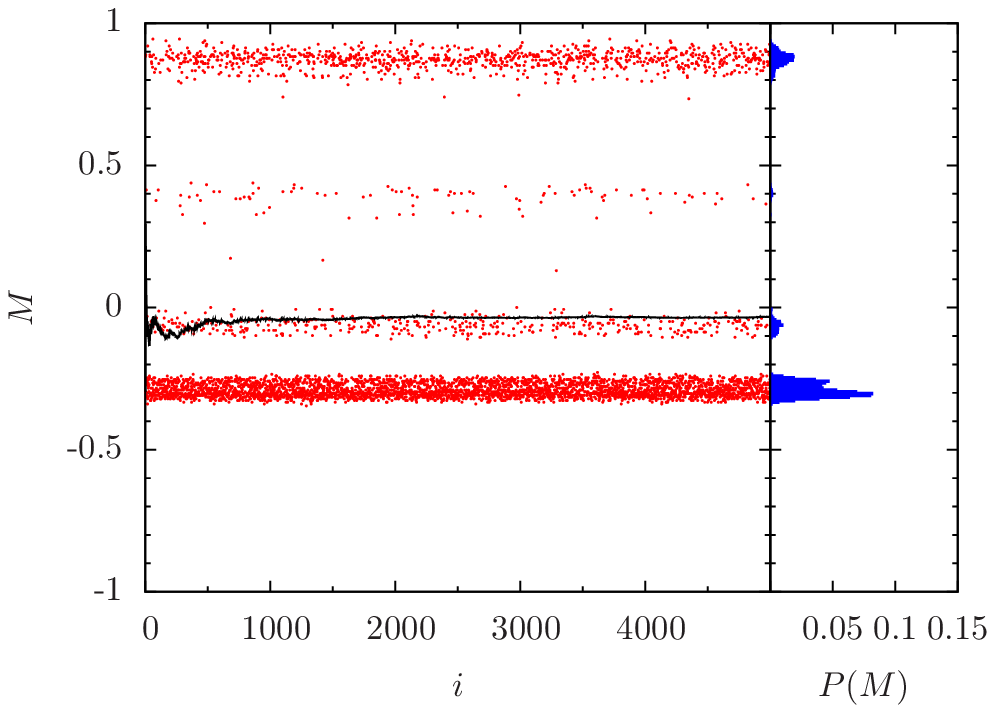}
\caption{(Color online) 
Left panels: Magnetization of ground-state configurations $M$ for
two independent samples (top/bottom panels) in the order they are
found (indexed by $i$) for a three-spin model
system with $18^2$ spins and $q=0.105$.
The solid line is a running average that converges toward zero,
suggesting that the ground-state magnetization is being sampled relatively 
fairly despite the slight bias shown in Fig.~\ref{fig:all_ground_states}.
Right panels: Histograms $P(M)$ of the magnetization.
}
\label{fig:magnetization}
\end{figure}

When the number of ground states is large, so that $N_g > N_p$, it also
appears that the subset of all ground states that the algorithm finds
is representative of the entire ensemble.  To test this, we study the
magnetization of the ground states. The results for two typical cases
are shown in Fig.~\ref{fig:magnetization}, where the magnetization
is shown for the first $5000$ ground states found in the order $i$
that they are found by the algorithm. There are no apparent temporal
correlations in the output and the running average quickly approaches
zero, although there are some statistical fluctuations about this
value because not all ground states have been found.  This, in turn,
justifies the use of this algorithm for sampling.

The two samples shown in Fig.~\ref{fig:magnetization} are for two
different instances of disorder with the same disorder strength
$q = 0.105$ and the same system size ($N = 18^2$).  This disorder
strength is near the ferromagnet to spin-glass phase transition
but slightly below it, so that the system is still ferromagnetic.
In the sample shown in the top panel, the states all resemble the pure
three-spin spin-glass problem, with $1/4$ of the configurations having
$M\approx1$, and the other $3/4$ having $M\approx-1/3$.  Unlike in
the pure case, there are many distinct ground states, but each one
follows the expectations for a ferromagnetic system. The sample
depicted in the bottom panel shows the onset of disorder: while the
majority of the states still have $M \approx 1$ or $M \approx -1/3$,
a sizable fraction of the states here have $M$ slightly below $0$, and
still others are between $0$ and $1/2$.  This implies that there are
large collective excitations with zero-energy, which is a signature of
glassy/frustrated systems. Thus, even for the same disorder strength,
the system will behave ferromagnetically for some instances of
the disorder and appear to be more glassy for others.

It is instructive to compare against the two-dimensional Ising
spin-glass model with $p=2$.  Even though efficient algorithms
have long been known for computing ground states in this
case,\cite{barahona:82} these techniques have not been directly
useful in finding all of the exponentially-many ground states,
or even in finding unbiased randomly chosen ground states.
Heuristic measures have therefore been employed to find the
ground-state magnetization,\cite{amoruso:04} although an efficient
algorithm for fair sampling of the ground states for this model
has been developed recently.\cite{thomas:09} This model has been
shown to exhibit reentrance from both zero- and finite-temperature
simulations,\cite{amoruso:04,parisen:09,thomas:11b} and we are not
aware of previous work on the three-spin model below the Nishimori
line for comparison.

\subsection{Ferromagnet--to--spin-glass transition}

\begin{table}
\caption{
Summary of simulation parameters for the two-dimensional three-spin
model of size $L_x \times L_y$, with population size $N_p$ and
averaging over $N_s$ different disorder instances. The smallest
[largest] value of $q$ simulated is $q_{\rm min}$ [$q_{\rm max}$]
in steps of $q_{\rm step}$.
\label{tab:3spin}
}
\begin{tabular*}{\columnwidth}{@{\extracolsep{\fill}} r r r r r r r }
\hline
\hline
$L_x$ & $L_y$ & $q_{\rm min}$ & $q_{\rm step}$ & $q_{\rm max}$ & $N_p$ & $N_s$ \\
\hline
 $6$ &  $6$ & $0.100$ & $0.001$ & $0.115$ &   $576$ & $10000$ \\
 $9$ &  $8$ & $0.100$ & $0.001$ & $0.115$ &  $1152$ & $10000$ \\
 $9$ & $10$ & $0.100$ & $0.001$ & $0.115$ &  $1440$ & $10000$ \\
$12$ & $12$ & $0.100$ & $0.001$ & $0.115$ &  $2304$ & $10000$ \\
$15$ & $14$ & $0.100$ & $0.001$ & $0.115$ &  $3360$ & $ 7000$ \\
$15$ & $16$ & $0.100$ & $0.001$ & $0.115$ &  $3840$ & $ 7000$ \\
$18$ & $18$ & $0.100$ & $0.001$ & $0.115$ &  $5184$ & $ 5000$ \\
$24$ & $24$ & $0.104$ & $0.001$ & $0.111$ &  $9216$ & $ 5000$ \\
$30$ & $30$ & $0.104$ & $0.001$ & $0.111$ & $14400$ & $ 5000$ \\
$36$ & $36$ & $0.104$ & $0.001$ & $0.111$ & $20736$ & $ 3000$ \\
\hline
\hline
\end{tabular*}
\end{table}

\begin{figure}[tb]
\includegraphics[width=\columnwidth]{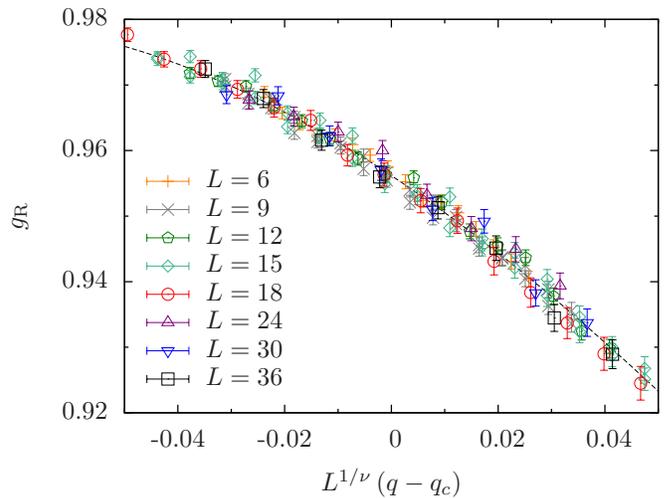}
\caption{(Color online) 
Scaling collapse of the restricted Binder ratio $g_{\rm R}$ according
to Eq.~(\ref{eq:scaling}) for the two-dimensional 3-spin Ising model
with system sizes $L=6$ -- $36$.  The dashed line is a curve fit to
a third-order polynomial. Optimal data collapse is obtained for
$q_c=0.1072(1)$ and $\nu=1.5(1)$.
}
\label{fig:scaling_collapse}
\end{figure}

Although this algorithm does not permit the computation of ground
states with high confidence in very large systems, the intermediate
system sizes accessible are adequate to precisely determine the
location of the $T=0$ ferromagnet--to--spin-glass phase transition as
the disorder strength $q$ varies. To do this, we compute the Binder
ratio\cite{binder:81} for a restricted set of magnetizations. Due to
the details of the four-fold degeneracy in this model including the
lack of a global spin-flip symmetry, the ferromagnetic state consists
of three states with weak negative magnetization for every highly
magnetized state with positive magnetization.  The paramagnetic
state has no sector with stronger magnetization than the others.
The fluctuations are strongest in the ferromagnetic sector, because
there is less change from the behaviors in the other sectors between
the two phases, so it is natural to focus on only the $1/4$ of states
that have a large magnetization. In Fig.~\ref{fig:magnetization},
these are the states with $M$ near $1$.  Let
\begin{equation}
M_0 = M \equiv \frac{1}{N}\sum_i S_i,
\end{equation}
and 
\begin{equation}
M_k = M \equiv \frac{1}{N}\sum_i S_ir_{ik}
\end{equation}
for numbered colors $k=1$, $2$, $3$ of the tripartite lattice, with
$r_{ik} = 1$ if site $i$ has color $k$ and $-1$ otherwise. We focus on
the state of the system with the largest magnetization, calling this
the restricted magnetization
\begin{equation}
M_R \equiv \mathrm{max}(M_0,M_1,M_2,M_3) .
\end{equation}
For any given spin configuration, it is straightforward to compute all
four of these magnetizations. This is analogous to taking the absolute
value for the $p=2$ Ising model, such that only configurations in
the same state are compared against one another. We then compute the
restricted Binder ratio $g_{\rm R}$ defined by
\begin{equation}
g_{\rm R} = \frac{1}{2}
	    \left( 
		3-\frac{[\langle M_R^4\rangle]_{\mathrm{av}}}{[\langle 
		M_R^2\rangle]^2_{\mathrm{av}}} 
	    \right),
\label{eq:binder_ratio}
\end{equation}
where $[\cdots]_{\rm av}$ is an average over samples and $\langle
\cdots \rangle$ represents an average over ground-state magnetizations
for the sample.  This is the same as the standard definition
of the Binder ratio except that the restricted magnetization $M_R$
is used in place of $M$.  For the $p=2$ Ising model, taking the
absolute value would not make any difference, but here the symmetric
states have different magnetizations.  The restricted Binder ratio 
is dimensionless and is $1$ in the ferromagnetic phase and $0$ in the
paramagnetic phase (here, the $T=0$ spin glass).  In a finite system
of linear scale $L$ with disorder strength in the vicinity of $q_c$,
$g_{\rm R}$ is expected to scale as
\begin{equation}
g_{\rm R} = \tilde{G}[ L^{1/\nu}(q-q_c)].
\label{eq:scaling}
\end{equation}
Equation (\ref{eq:scaling}) allows for the extraction of the critical
disorder strength $q_c$ as well as the correlation-length exponent
$\nu$. To estimate the optimal values of the critical parameters we
perform a finite-size scaling analysis where we tune $q_c$ and $\nu$
until the chi-squared of a fit to a third-order polynomial in the
vicinity of $L^{1/\nu}(q - q_c) \lesssim 1$ is minimized. Error bars
are determined by a bootstrap analysis.\cite{efron:79,katzgraber:06} Our best
estimates of the critical parameters are
\begin{equation}
q_c = 0.1072(1)
\;\;\;\;\;\;\;\;\;\;\;\;
\nu = 1.5(1) \, .
\end{equation}
The scaling collapse using these critical parameters is shown in
Fig.~\ref{fig:scaling_collapse}.  To verify our results, we have also
computed the standard Binder ratio on the unrestricted magnetizations
from the same data, which produced consistent results for both $q_c$
and $\nu$.  The only noticeable difference in the results is the
form of the scaling function, that is, $g_{\rm R}(q=q_c) \approx 0.955$,
while $g(q=q_c) \approx 0.23$.
A summary of the simulation parameters is shown in Table~\ref{tab:3spin}.

\begin{figure}[tb]
\includegraphics[width=\columnwidth]{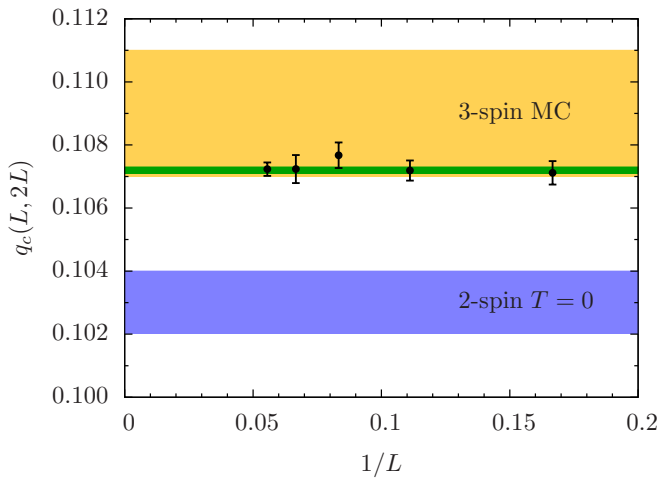}
\caption{(Color online) 
Measured values of $q_c$ as a function of system size.  Each data
point is the value of $q_c$ extracted from the scaling collapse
of a system of size $L$ with a system of size $2L$, for $L=6$,
$9$, $12$, $15$, and $18$.  The yellow (light) region is the
estimate $q_c = 0.109(2)$ from Ref.~\onlinecite{katzgraber:09c} from
finite-temperature Monte Carlo (MC) simulations of the 3-spin model. 
The solid green (darker) line
is the $T=0$ estimate $q_c=0.1072(1)$ in this study, and the blue (dark)
region is, for comparison, the $T=0$ estimate $q_c=0.103(1)$
for the two-dimensional 2-spin
Ising model from Ref.~\onlinecite{amoruso:04}.  The fact that the
data (solid green line) and the result for the two-dimensional Ising
model (blue/dark region) do not overlap clearly shows that while both
models apparently share the same multicritical point, the behavior
for temperatures {\em below} the multicritical point is quite different.
}
\label{fig:scaling_plot}
\end{figure}

Finally, to treat the possibility of finite-size corrections
to scaling, we separate the components of the fit in
Fig.~\ref{fig:scaling_collapse} into $L$--$2L$ pairs.  We perform the
curve-fitting procedure to extract $q_c$ using each pair of system
sizes.  The triangular lattice is tripartite only for $L_x$~mod~$3 =
0$ and $L_y$~mod~$2 = 0$, so these comparisons are only possible for
and $N = L\times L$ triangular lattice if $L$ is a multiple of $6$.
To increase the number of cases we can handle, we consider $L_x$
to be any multiple of $3$, and if $L_x$ is not divisible by $2$,
we simulate for both $L_x\times (L_x-1)$ and $L_x\times (L_x+1)$
systems.  This gives two results which are presumably biased in
opposite directions, which can be seen for the odd system sizes in
Fig.~\ref{fig:scaling_collapse}.  However, our statistical errors are
as large as the bias introduced, so we simply average the two results.
In comparing sequences of $L$--$2L$ pairs, we check if the result
depends on $L$ to extrapolate to the thermodynamic limit.  We find
that the extracted values of $q_c(L,2L)$ show no visible dependence on
system size, as shown in Fig.~\ref{fig:scaling_plot}.  Thus the above
estimates of $q_c$ and $\nu$ appear to be indicative of the values in
the thermodynamic limit.  In contrast to the two-dimensional Ising
model, the three-spin model appears to have much weaker reentrance,
if it is reentrant at all.  The phase diagrams for the two models
therefore differ considerably {\em below} the Nishimori line.

\section{Three-dimensional Ising spin glass}
\label{sec:3d}

As another application of this algorithm, we investigate the
ferromagnet--to--spin-glass transition in the three-dimensional Edwards
Anderson Ising spin glass with bimodal disorder. This model is
important in the study of disordered materials.\cite{binder:86}
It is also notoriously difficult to study numerically.  It is
NP-hard,\cite{barahona:82} so exact methods are quite limited in
scope,\cite{liers:04} and even highly-sophisticated heuristic methods
can only produce reliable results up to $L=14$.\cite{hartmann:99b}

\subsection{Model}

The two-spin Edwards-Anderson Ising spin glass is given by the Hamiltonian
\begin{eqnarray}
\mathcal{H} & = & \sum_{\langle i j \rangle} J_{ij} S_i S_j,
\end{eqnarray}
where the spins $S_i$ sit at the sites of a cubic lattice and pairwise
interactions ($p = 2$) are over nearest neighbors.  Here we consider
bimodal disorder, where the bond strengths are given by the same
probability distribution as the plaquettes in Eq.~(\ref{eq:bimodal}),
\begin{equation}
P(J_{i j}) =      (1-q)\delta(J_{i j}-1)
                                +q\delta(J_{i j}+1).
\end{equation}

\begin{figure}[tb]
\includegraphics[width=\columnwidth]{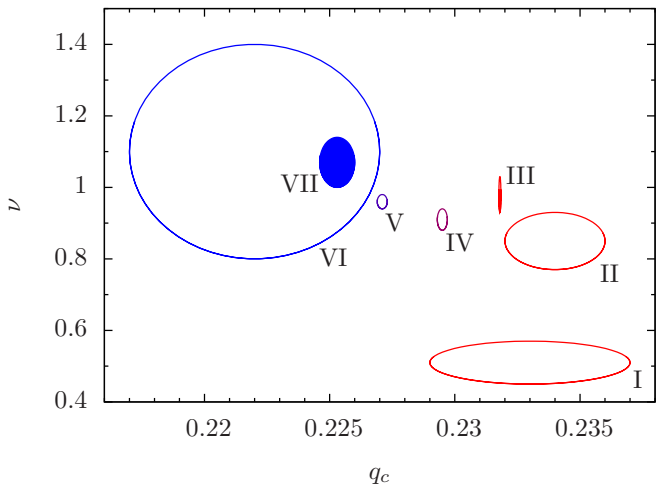}
\caption{(Color online) 
Graphical representation of different estimates of the
critical concentration $q_c$ and the critical exponent $\nu$
for the three-dimensional Edwards-Anderson model, as shown in
Table~\ref{tab:literature_summary} (tagged via roman numerals). The
center of an ellipse corresponds to the optimal estimate of $q_c$
and $\nu$, while the size of the ellipse corresponds to the error bar
attached to it.  The red ellipses [I -- III] are at the multicritical
point, which is believed be in a different universality class than
the low- and zero-temperature case.  The blue ellipses [VI, VII]
are zero-temperature estimates (the filled ellipse represents the
results from this study) and the purple ellipses [IV, V] are for finite
temperatures {\em below} the multicritical point.  The fact that the
finite-temperature (red) and zero-temperature (blue) estimates do
not overlap in the horizontal direction is indicative of a reentrant
behavior in the phase diagram of the model.
}
\label{fig:literature_summary}
\end{figure}

The zero-temperature ferromagnet--to--spin-glass transition has been
investigated for this problem using a sophisticated heuristic
algorithm,\cite{hartmann:99b} where the critical disorder strength was
found to be $q_c = 0.222(5)$.  This work did not take into account possible
biases in the sampling of ground states, which could lead to incorrect results,
as pointed out in Ref.~\onlinecite{sandvik:99}.  The finite-temperature multicritical
point has been evaluated for this model in a number of studies,
with the results summarized in Table~\ref{tab:literature_summary}
and shown graphically in Fig.~\ref{fig:literature_summary}.
Note that, where necessary, we have converted estimates so they are
all in the same terms.  In particular, Ref.~\onlinecite{singh:91a}
quotes only the temperature $T_c^*$ of the multicritical point,
which is easily converted to $q_c^*$, and the value given for $\nu$
in Ref.~\onlinecite{hasenbusch:07} is for thermal perturbations, with
$\nu=1/y_2$, while the value we quote ($1/y_1$) is more relevant to low-temperature studies because
it describes the disorder
perturbations that are used at the zero-temperature fixed point.
The temperature of the multicritical point
(MCP) can be determined directly from the Nishimori condition
\begin{eqnarray} 
1-2 q &=& \tanh (1/T), 
\end{eqnarray} 
with values in the literature in the range $T_c^*\approx1.67$---$1.69$.
It is believed that the data below the multicritical point are
universal and in a different universality class than {\em at} the
multicritical point.\cite{ceccarelli:11}  This view is consistent with
the known data.  Furthermore, the $q_c$ values show strong evidence
for reentrance in the phase diagram.

\begin{table}
\caption{
Selection of different estimates of the critical concentration $q_c$
and the critical exponent $\nu$ for the three-dimensional bimodal
Edwards-Anderson Ising spin glass.  The estimates are plotted in
Fig.~\ref{fig:literature_summary} and show some variations.
Note that the values quoted for $\nu$ at the multicritical point (MCP)
correspond to disorder perturbations and are not the same $\nu$ one
would measure by only perturbing the temperature $T$.\cite{hasenbusch:07}
(The roman numerals are used to tag the different estimates in the figure).
\label{tab:literature_summary}
}
\begin{tabular*}{\columnwidth}{@{\extracolsep{\fill}} l l l l }
\hline
\hline
Authors & $T$ & $q_c$ & $\nu$ \\
\hline
Ozeki and Nishimori [Ref.~\onlinecite{ozeki:87}, I]                 & MCP & 0.233(4) & 0.51(6) \\
Singh [Ref.~\onlinecite{singh:91a}, II]                             & MCP & 0.234(2) & 0.85(8) \\
Hasenbusch {\textit et.~al.} [Ref.~\onlinecite{hasenbusch:07}, III] & MCP & 0.23180(4) & 0.98(5)\\
Ceccarelli {\textit et.~al.} [Ref.~\onlinecite{ceccarelli:11}, IV]  & 1.0 & 0.2295(2) & 0.91(3) \\
Ceccarelli {\textit et.~al.} [Ref.~\onlinecite{ceccarelli:11}, V]   & 0.5 & 0.2271(2) & 0.96(2) \\
Hartmann [Ref.~\onlinecite{hartmann:99b}, VI]                       & 0.0 & 0.222(5) & 1.1(3) \\
\hline
This study [VII]                                                    & 0.0 & 0.2253(7) & 1.07(7) \\
\hline
\hline
\end{tabular*}
\end{table}

\subsection{Results}

\begin{table}
\caption{
Summary of simulation parameters for the three-dimensional
Edwards-Anderson model of size $L^3$, with population size $N_p$ and
averaging over $N_s$ different instances of disorder.  The smallest
[largest] value of $q$ simulated is $q_{\rm min}$ [$q_{\rm max}$]
in steps of $q_{\rm step}$.
\label{tab:3DISG}
}
\begin{tabular*}{\columnwidth}{@{\extracolsep{\fill}} r r r r r r}
\hline
\hline
$L$ & $q_{\rm min}$ & $q_{\rm step}$ & $q_{\rm max}$ & $N_p$ & $N_s$ \\
\hline
 $4$ & $0.20$ & $0.01$ & $0.25$ &  $1024$ & $3000$ \\
 $6$ & $0.20$ & $0.01$ & $0.25$ &  $3456$ & $3000$ \\
 $8$ & $0.21$ & $0.01$ & $0.24$ &  $8192$ & $1700$ \\
$10$ & $0.21$ & $0.01$ & $0.24$ & $16000$ & $1200$ \\
\hline
\hline
\end{tabular*}
\end{table}

\begin{figure}[tb]
\includegraphics[width=\columnwidth]{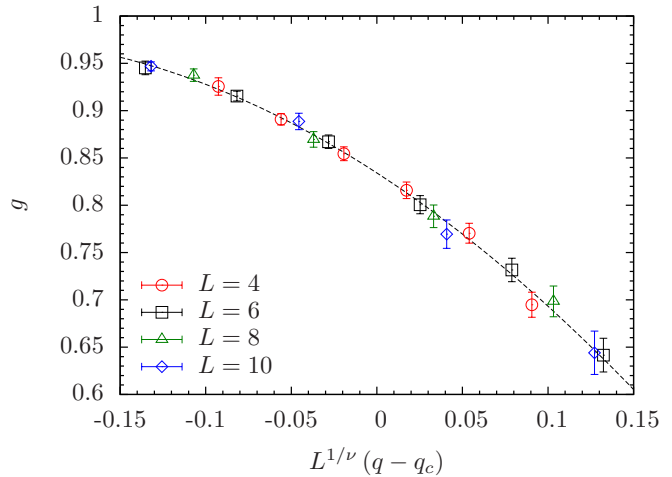}
\caption{(Color online) 
Scaling collapse of the Binder ratio $g$ according to
Eq.~(\ref{eq:scaling}) for the three-dimensional Edwards
Anderson model.  The dashed line is a curve fit to a third-order
polynomial. Optimal data collapse is obtained for $q_c=0.2253(7)$
and $\nu=1.07(7)$.
}
\label{fig:binder3d}
\end{figure}

We use the {\em same code} for simulating the three-spin Ising spin
glass to also find ground states of the $p=2$ Ising spin glass in
three dimensions.  We have computed the Binder ratio in the vicinity
of the critical disorder strength for systems of sizes $4^3$--$10^3$.
In the Edwards-Anderson model, the restricted Binder ratio is identical
to the standard definition of the Binder ratio because only even
moments of the magnetization are used; here $g = g_\mathrm{R}$, so the
scaling relation in Eq.~(\ref{eq:scaling}) is again expected to hold.
Figure \ref{fig:binder3d} shows the scaling collapse for this model.
The optimal values of the critical parameters extracted from this
best fit are
\begin{equation}
q_c = 0.2253(7)
\;\;\;\;\;\;\;\;\;\;\;\;
\nu = 1.07(7) , 
\end{equation}
in agreement with the results of Ref.~\onlinecite{hartmann:99b}.
The system sizes available do not permit the more thorough finite-size
scaling analysis shown in Fig.~\ref{fig:scaling_plot}, although the
corrections to scaling appear to be quite weak, suggesting that this
result is robust.
A summary of the simulation parameters is shown in Table~\ref{tab:3DISG}.

\section{Conclusions}

We have demonstrated the applicability of the genetic ground-state
algorithm presented in Ref.~\onlinecite{thomas:11} to sample among the
many ground states in highly-degenerate NP-hard models.  Using this
technique, spin configurations and therefore magnetizations of
typical ground-state configurations may be accessed, allowing the
construction of parameters such as the Binder ratio without bias,
that are effective for determining the location of phase transitions.

Above the Nishimori line, the phase diagram of the three-spin Ising
model on a triangular lattice closely resembles that of the
square-lattice Ising model with two-spin interactions.  Although the
models are in different universality classes, the measured critical
values for the pure case and on the Nishimori line are the same.  The
results here show that this similarity does not extend {\em below} the
Nishimori line, that is, the models have quite different transition
disorder strengths at $T = 0$ (see Fig.~\ref{fig:scaling_plot}).  We
show that $q_c(T = 0) = 0.1072(1)$.

Furthermore, using a modest numerical effort we determine the
ferromagnet--to--spin-glass transition for the three-dimensional
Edwards-Anderson Ising spin glass, that is, $q_c=0.2253(7)$, in agreement
with previous measurements.\cite{hartmann:99b}  Comparing against
estimates of the multicritical point, this three-dimensional model
possesses reentrance in its phase diagram by an amount which is similar
to the two-dimensional model (similar results were obtained recently
in Ref.~\onlinecite{ceccarelli:11}).

\begin{acknowledgments} 

We would like to thank Alexander K.~Hartmann, Andrea Pelissetto, and
Ettore Vicari for useful discussions.  H.G.K.~acknowledges support from
the SNF (Grant No.~PP002-114713).  The authors acknowledge the Texas
Advanced Computing Center (TACC) at The University of Texas at Austin
for providing HPC resources (Ranger Sun Constellation Linux Cluster) and
ETH Zurich for CPU time on the Brutus cluster.

\end{acknowledgments}

\bibliography{refs,comments}

\end{document}